\documentclass[aps,onecolumn,groupedaddress,floatfix]{revtex4-1}
\usepackage{amssymb,amsmath}
\usepackage{graphicx}
\usepackage{subfigure}
\usepackage[english]{babel}
\usepackage{float}
\usepackage{color}

\usepackage{lipsum}
\begin{document}
\newcommand{\be}{\begin{equation}}
\newcommand{\ee}{\end{equation}}
\newcommand{\rojo}[1]{\textcolor{red}{#1}}

\makeatletter
\DeclareRobustCommand*\cal{\@fontswitch\relax\mathcal}
\makeatother

\title{The fractional nonlinear $\cal{PT}$ dimer}

\author{Mario I. Molina}
\affiliation{Departamento de F\'{\i}sica, Facultad de Ciencias, Universidad de Chile, Casilla 653, Santiago, Chile}

\date{\today }

\begin{abstract} 
We examine a fractional Discrete Nonlinear Schrodinger dimer, where the usual first-order derivative of the time evolution is replaced by a non integer-order derivative. 
The dimer is nonlinear (Kerr) and ${\cal{PT}}$-symmetric, and we examine the exchange dynamics between both sites. By means of the Laplace transformation technique, the linear ${\cal{PT}}$ dimer is solved in closed form in terms of Mittag-Leffler functions, while for the nonlinear regime, we resort to numerical computations using the direct explicit Grunwald algorithm. In general, the main effect of the fractional derivative is the onset of a monotonically decreasing time envelope for the amplitude of the oscillatory exchange. In the presence of ${\cal{PT}}$ symmetry, the dynamics shows damped oscillations for small gain/loss in both sites, while at higher gain/loss parameter values, the amplitudes of both sites grows unbounded. In the presence of nonlinearity, selftrapping is still possible although the trapped fraction decreases as the nonlinearity is increased past threshold, in marked contrast with the standard case.
\end{abstract}

\maketitle

\section{Introduction}

The topic of fractional calculus has experienced a rekindled interest in recent times. Essentially, it extends the notion of a derivative or an integral of integer order, to one of a fractional order, $(d^n/dx^n)\rightarrow (d^\alpha/d x^\alpha)$ for real $\alpha$. The subject has a long history, dating back to letters exchanged between Leibnitz and L'Hopital, and later contributions by Euler, Laplace, Riemann, Liouville, and Caputo to name some\cite{hilfer,west,podlubnik,herrmann,Caputo}.  The starting point was the computation of $d^\alpha x^{k}/d x^\alpha$, where $\alpha$ is a non-integer number: 
\be 
{d^{n} x^k\over{d x^n}}= {\Gamma(k+1)\over{\Gamma(k-n+1)}} x^{k-n} \rightarrow {d^\alpha x^k\over{d x^\alpha}} = {\Gamma(k+1)\over{\Gamma(k-\alpha+1)}} x^{k-\alpha}.\label{eq1}
\ee
For instance, $(d^{1/2}/d x^{1/2}) x = (2/\sqrt{\pi}) \sqrt{x}$, and $d x/d x = (d^{1/2}/d x^{1/2}) (d^{1/2}/d x^{1/2}) x = (2\sqrt{\pi})(\Gamma(3)/\Gamma(1)) x^{0} = 1$, as expected.
From Eq.(\ref{eq1}) the fractional derivative of an analytic function $f(x)=\sum_{k} a_{k} x^{k}$ can be computed by deriving term by term. This basic procedure is not exempt from ambiguities. For instance, $(d^\alpha/d x^{\alpha})\ 1=(d^\alpha x^{0}/d x^\alpha)=(1/\Gamma(1-\alpha)) x^{-\alpha}\neq 0$, according to Eq.(\ref{eq1}). However, one could have also taken $(d^{\alpha-1}/d x^{\alpha-1})(d/dx)\ 1=0$. For the case of a fractional integral, a more rigorous  starting point is Cauchy's formula for the integral of a function. From the definition
\be
I_{x}^{1} f(x) = \int_{0}^{x} f(s) d s, \label{eq2}
\ee
we apply the Laplace transform $\cal{L}$ to both sides of Eq.(\ref{eq2})
\be
{\cal{L}}\ \{I_{x}^{1}\ f(x)\}= (1/s)\ {\cal{L}}\{f(x)\}. 
\ee
After $n$ integrations, one obtains
\be
{\cal{L}}\ \{I_{x}^{n}\ f(x)\}= (1/s^n)\ {\cal{L}}\{f(x)\}.
\ee
Extension to fractional $\alpha$ is direct:
\be
{\cal{L}}\ \{I_{x}^{\alpha}\ f(x)\}= (1/s^\alpha)\ {\cal{L}}\{f(x)\}.
\label{eq3}
\ee
After noting that the RHS of Eq.(\ref{eq3}) is the product of two Laplace transforms we have, after using the convolution theorem
\be
I_{x}^{\alpha}\ f(x) = {1\over{\Gamma(\alpha)}} \int_{0}^{x} {f(s)\over{(x-s)^{1-\alpha}}}\ ds. \label{eq4}
\ee
From this definition, it is possible to define the fractional derivative of a function $f(x)$ as 
\begin{eqnarray}
& & {d^\alpha f(x)\over{d x^\alpha}} = \left({d^m \over{d x^m}}\right)\ I_{x}^{m-\alpha}\ f(x) = \nonumber\\
& & = {d^{m}\over{d x^m}} \left[ {1\over{\Gamma(m-\alpha)}} \int_{0}^{x} (x-s)^{m-\alpha-1} f(s)\ ds\ \right],\label{eq5}
\end{eqnarray}
where, $m-1<\alpha<m$. Eq.(\ref{eq5}) is known as the Riemann-Liouville form. An alternative, closely related form, is the Caputo formula\cite{Caputo}:
\begin{eqnarray}
& & {d^\alpha f(x)\over{d x^\alpha}} = I_{x}^{m-\alpha}\left({d^{m}\over{d x^m}}\right)f(x) = \nonumber\\ & &= {1\over{\Gamma(m-\alpha)}} \int_{0}^{x}(x-s)^{m-\alpha-1} f^{(m)}(s) ds,
\end{eqnarray}
which has some advantages over the Riemann-Liouvuille form for differential equations with initial values. 
The various technical matters that arise in fractional calculus have prompted a whole line of research that has extended to current times.
Long regarded as a mathematical curiosity, it has now regained interest due to its potential applications to complex problems in several fields: fluid mechanics\cite{caffarelli, constantin}, fractional kinetics and anomalous diffusion\cite{metzler, sokolov, zaslavsky}, strange kinetics\cite{shlesinger}, fractional quantum mechanics\cite{laskin1,laskin2}, Levy processes in quantum mechanics\cite{petroni}, plasmas\cite{allen}, electrical propagation in cardiac tissue\cite{bueno} and biological invasions\cite{berestycki}. In general, fractional calculus constitutes a natural formalism for the description of memory and non-locality effects found in various complex systems. Experimental realizations are not straightforward given the nonlocal character of the coupling, however some optical setups have been suggested that could measure the effect of fractionality on new beam solutions and new optical devices\cite{longhi,yiqi}

On the other hand, when dealing with effectively discrete, interacting units, as one encounters in atomic physics (interacting atoms), or in optics ( coupled optical fibers), it is common to deal with discrete versions of the continuum Schr\"{o}dinger equation, or the paraxial wave equation. The effective discreteness comes from expanding the solution sought in terms of (continuous) modes that can be labelled unambiguously.  The simplest of such examples is the bonding, anti-bonding electronic mode that one finds for a two-sites (dimer) molecule after diagonalizing the two-site Schr\"{o}dinger equation in the tight-binding approach. Something similar happens in optics, where the paraxial equation is formally equivalent to the Schr\"{o}dnger equation. In that case, for two optical waveguides, the total electric field is expanded in terms of the electromagnetic modes in each guide which interact through the  evanescent field between the two guides giving rise to a transversal dynamics for the optical power. The procedure can be extended to $N$ interacting units, either atoms or waveguides, where the relevant dynamics is given by a discretized version of the Schr\"{o}dinger equation for $N$ units\cite{saleh,yeh}. Of course, at the end one has to collect all the discrete amplitudes and multiply them by the corresponding  continuous mode profiles and superpose them, to obtain the final field. The simplest case $N=2$ is termed a dimer and oftentimes constitute a basic starting point when studying an interacting, discrete system. Ensembles of interacting dimers have been studied before in classical and quantum statistics\cite{kasteleyn,moessner,misguich}, and more recently, they have been considered in model of correlated disorder\cite{szameit} and in magnetic metamaterial  modeling\cite{MM}.

In this work we consider the discrete Schr\"{o}dinger equation for a dimer system, where the standard time derivative is replaced by a fractional one.
The dimer considered is rather general and contains asymmetry, $\cal{PT}$ symmetry and nonlinearity (Fig.1). Our main interest is in ascertaining the effect of the fractional derivative on the excitation exchange between the sites, its stability and selftrapping behavior, for several cases of interest.

\section{The fractional dimer} 

Let us consider the fractional evolution equations for a general  nonlinear dimer
\begin{eqnarray}
i {d^{\alpha} C_{1}(t)\over{d t^{\alpha}}} + \epsilon_{1} C_{1}(t) + V C_{2}(t) + \chi |C_{1}(t)|^2 C_{1}(t)&=&0\nonumber\\
i {d^{\alpha} C_{2}(t)\over{d t^{\alpha}}} + \epsilon_{2} C_{2}(t) + V C_{1}(t) + \chi |C_{2}(t)|^2 C_{2}(t)&=&0\ \ \ \ \label{eq9}
\end{eqnarray}
where $\alpha$ is the fractional order of the (Caputo) derivatives with $0<\alpha<1$. Quantities $C_{1,2}$ are probability amplitudes in a quantum context, or electric field amplitudes, in an optical setting. Parameter $V$ is the coupling term and $\chi$ is the nonlinearity parameter.
\begin{figure}[t]
 \includegraphics[scale=0.15]{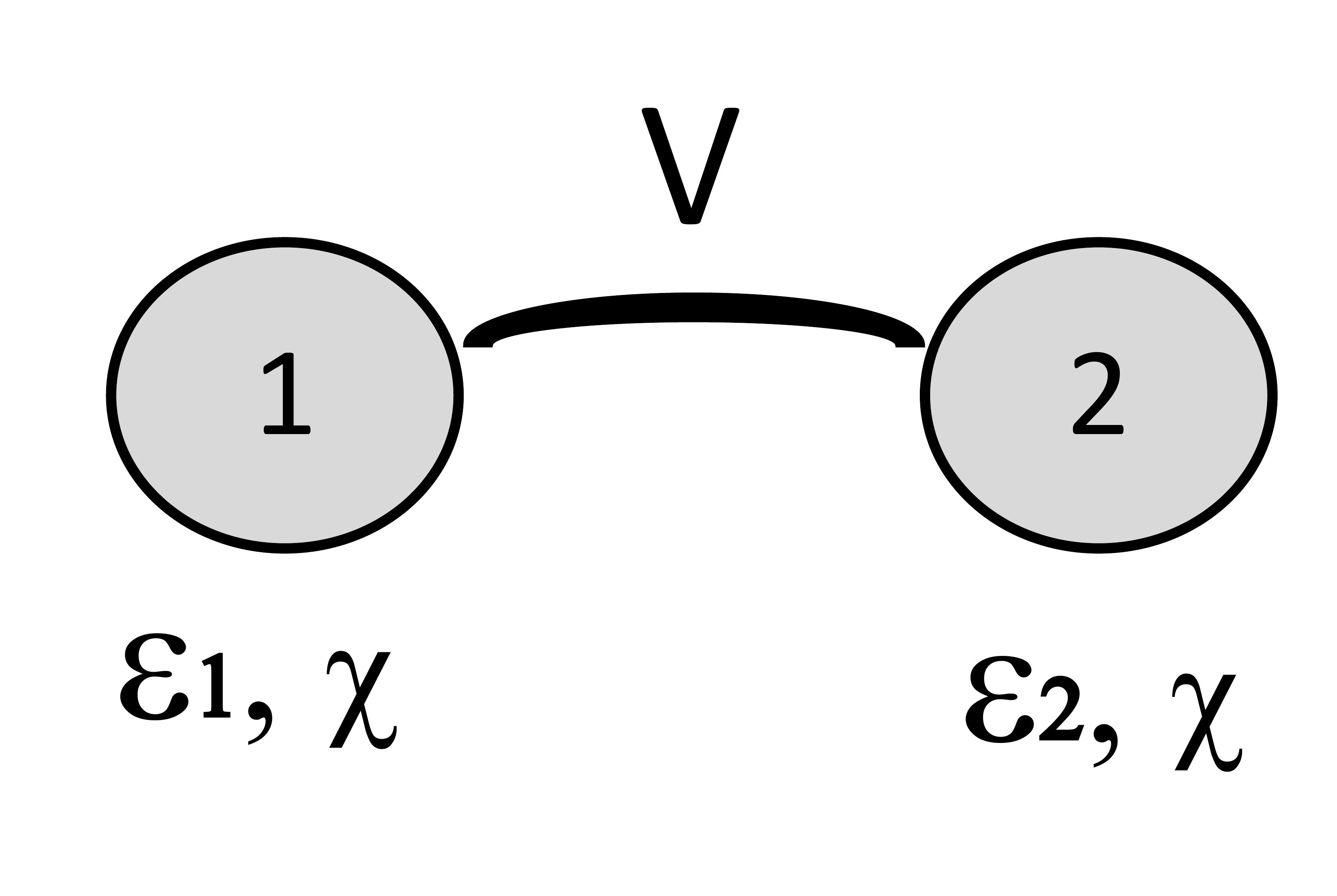}
  \caption{ Nonlinear anisotropic fractional dimer where the excitation is on site $1$ initially ($t=0$). In the $\cal{PT}$ case, $\epsilon_{1}=-\epsilon_{2}=i \gamma$
  }
  \label{fig1}
\end{figure}

Let us first consider the case of a general linear ($\chi=0$) dimer, and assume $C_{1}(0)=1, C_{2}(0)=0$. We will solve this case in closed form by the use of Laplace transforms: For $0<\alpha<1$, the Laplace transform of the Caputo fractional derivative of order $\alpha$ is given by
\be
{\cal{L}}\{f(t)\} = s^{\alpha} {\cal{L}}\{f(t)\} - s^{\alpha-1} f(0^{+}).
\ee
After applying the Laplace transform $\cal{L}$ to both sides of 
Eq.(\ref{eq9}) we have
\begin{eqnarray}
i (s^{\alpha} {\cal{L}}(C_{1}) - s^{\alpha-1}) + \epsilon_{1} {\cal{L}}(C_{1})+ V {\cal{L}}(C_{2})&=&0\nonumber\\
i\ s^{\alpha} {\cal{L}}(C_{2}) + \epsilon_{2} {\cal{L}}(C_{2})+ V {\cal{L}}(C_{1})&=&0.
\end{eqnarray}
Solving for ${\cal{L}}(C_{1})$ and ${\cal{L}}(C_{2})$ gives:
\be
{\cal{L}}(C_{1}) = {- i (\epsilon_{2} + i s^{\alpha}) s^{\alpha-1}\over{s^{2 \alpha} - i (\epsilon_{1}+\epsilon_{2})s^\alpha + V^2-\epsilon_{1} \epsilon_{2}}}
\ee
and
\be
{\cal{L}}(C_{2}) = {i\ V s^{\alpha-1}\over{s^{2 \alpha} - i (\epsilon_{1} + \epsilon_{2}) s^\alpha + V^2-\epsilon_{1} \epsilon_{2}}}.
\ee
Using the inverse Laplace formula\cite{haubold}
\begin{eqnarray}
& &{\cal{L}}^{-1}\left\{ {s^{\rho-1}\over{s^\alpha+a s^\beta+b}}\right\} = \nonumber\\
& & t^{\alpha-\rho}\sum_{r=0}^{\infty} (-a)^r t^{(\alpha-\beta)r} E_{\alpha,\alpha+(\alpha-\beta)r-\rho+1}^{r+1} (-b t^\alpha),\ \ \ \ 
\end{eqnarray}
we obtain:
\begin{eqnarray}
& & C_{1}(t) = \sum_{r=0}^{\infty} (i (\epsilon_{1}+\epsilon_{2}))^{r}\ t^{\alpha r} E_{2\alpha,\alpha r +1}^{r+1} ((\epsilon_{1}\epsilon_{2}-V^2)t^{2 \alpha})\nonumber\\
& & -i \epsilon_{2} t^{\alpha}\sum_{r=0}^{\infty} (i (\epsilon_{1}+\epsilon_{2}))^r t^{\alpha r} E_{2\alpha,\alpha (1+r)+1}^{r+1}((\epsilon_{1}\epsilon_{2}-V^2) t^{2\alpha})\ \ \ \ \ \ \ \ \ \label{eq15} 
\end{eqnarray}
\be
C_{2}(t)=i V t^\alpha \sum_{r=0}^{\infty} (i (\epsilon_{1}+\epsilon_{2}))^r t^{\alpha r} E_{2\alpha,\alpha(1+r)+1}^{r+1} ((\epsilon_{1}\epsilon_{2}-V^2) t^{2\alpha})\label{eq16}
\ee
where $E_{\alpha,\beta}^{\gamma}(z)$ is defined as
\be
E_{\alpha,\beta}^{\gamma}(z) = \sum_{k=0}^{\infty} {(\gamma)_{k}\ z^{k}\over{k!\ \Gamma(\alpha k + \beta)}} 
\ee
where $(\gamma)_{n} = \Gamma(\gamma+n)/\Gamma(\gamma)$, 
and  
$\alpha, \beta, \gamma \in {\cal{C}}$, $\mbox{Re}(\alpha)>0, \mbox{Re}(\beta)>0, z\in C $. 
\begin{figure}[t]
 \includegraphics[scale=0.3]{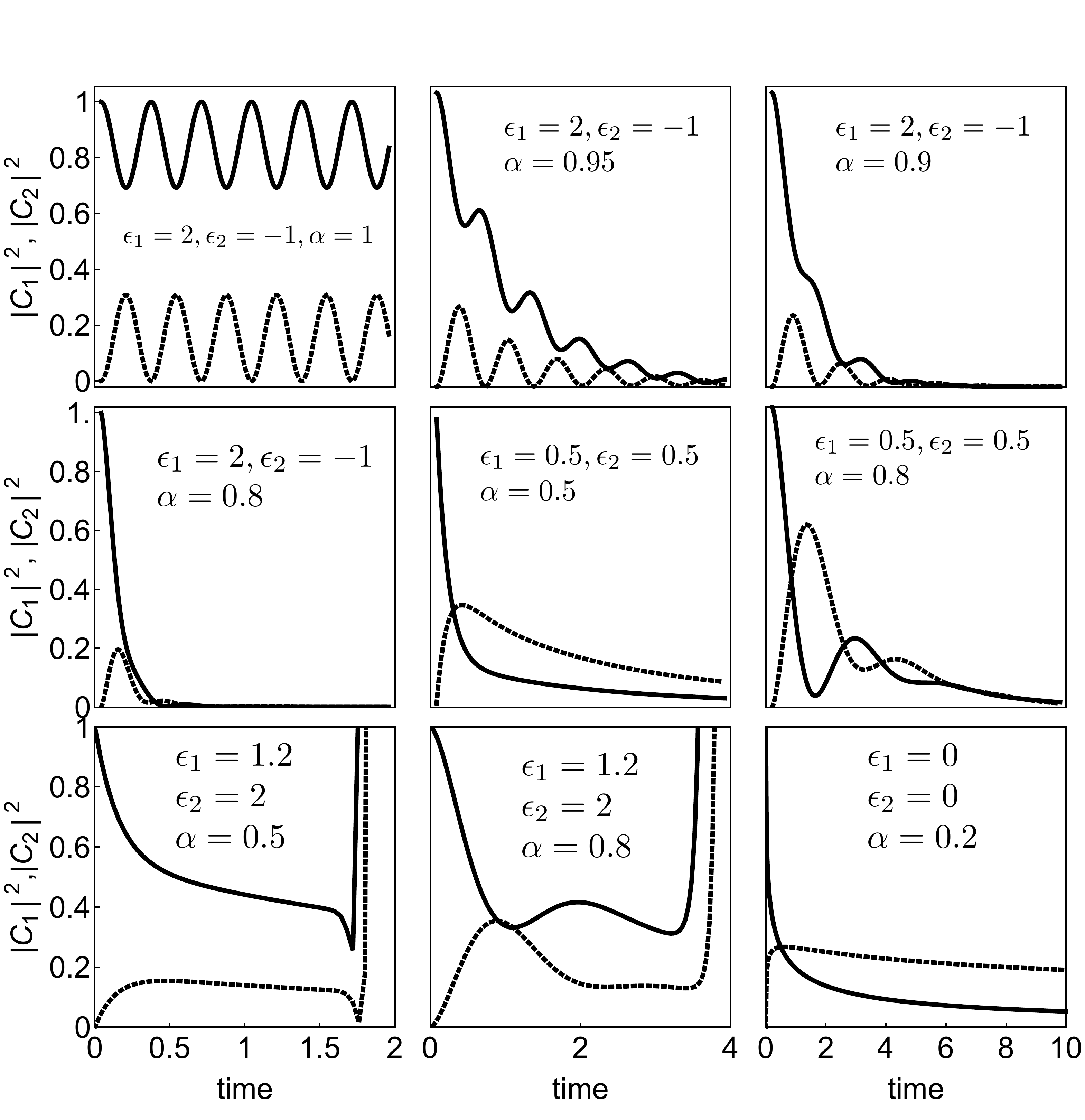}
  \caption{ Dimer amplitudes for the linear case ($\chi=0$) and several site energy parameters $\epsilon_{1}, \epsilon_{2}$ and different fractional derivative orders $\alpha$. Solid(dashed) line denotes $|C_{1}|^2$ 
  $(|C_{2}|^2)$.
  }
  \label{fig2}
\end{figure}
Figure 2 shows examples of the time evolution of the square of the dimer amplitudes, for several site energy parameters, and fractional derivative orders. In general we observe that, as soon as $\alpha$ differs from unity, the dynamics is either bounded or unbounded, depending on the values of the site energy parameters. For the bounded cases, there is some oscillation initially, with a decreasing envelope towards zero.

\subsection{The linear ${\cal{PT}}$ dimer}

A particularly interesting case of Eq.(\ref{eq9}), is 
the fractional ${\cal{PT}}$-symmetric  dimer. For systems that are invariant under the combined operations of parity (${\cal{P}}$) and time reversal (${\cal{T}}$), it was shown that they display a real eigenvalue spectrum, even though the underlying Hamiltonian is not hermitian\cite{PT_theory1,PT_theory2}. In these systems there is a balance between gain and loss, leading to a bounded dynamics. However, as the gain/loss parameter exceeds a certain value the system undergoes a spontaneous symmetry breaking, where two or more eigenvalues become complex. At that point, the system loses its balance and its dynamics becomes unbounded. 
According to the general theory, for our system to be ${\cal{PT}}$ symmetric, the real part of the site energies in Eq.(\ref{eq9}) must be even in space while the imaginary part must be odd: $\mbox{Re}(\epsilon_{1})=\mbox{Re}(\epsilon_{2})$ and $\mbox{Im}(\epsilon_{1})=-\mbox{Im}(\epsilon_{2})$. For simplicity we take the real parts of
$\epsilon_{1}$, $\epsilon_{2}$ as zero and thus, $\epsilon_{1}=-\epsilon_{2}\equiv i\ \epsilon$, where $\epsilon$ is the gain/loss parameter. This leave us with the equations:
\begin{eqnarray}
i {d^{\alpha} C_{1}(t)\over{d t^{\alpha}}} + i \epsilon C_{1}(t) + V C_{2}(t)&=&0\nonumber\\
i {d^{\alpha} C_{2}(t)\over{d t^{\alpha}}} - i \epsilon C_{2}(t) + V C_{1}(t)&=&0\ \ \ \ \label{eq20}
\end{eqnarray}
whose exact solutions can be extracted from the general solution, Eqs.(\ref{eq15}),(\ref{eq16}) as
\begin{eqnarray}
C_{1}(t)&=&-\epsilon t^{\alpha} E_{2 \alpha,\alpha+1}((\epsilon^2-V^2)t^{2 \alpha}) +
E_{2 \alpha,1}((\epsilon^2-V^2)t^{2 \alpha})\nonumber\\
C_{2}(t)&=& i V t^{\alpha} E_{2 \alpha,\alpha+1}((\epsilon^2-V^2) t^{2 \alpha}),
\label{eq21}
\end{eqnarray}
where, $E_{\alpha,\beta}(z)=E_{\alpha,\beta}^{1}(z)$ is known as the generalized Mittag-Leffler function
\be
E_{\alpha, \beta}(z) = \sum_{k} {z^{k}\over{\Gamma(\alpha k + \beta)}}.
\ee
It is the natural extension of the exponential function and plays the same rol for fractional differential equations, as the exponential function does for the standard integer differential equations. Figure 3 shows examples of time evolutions for $|C_{1}|^2, |C_{2}|^2$ for several fractional orders and several gain/loss parameter values.
\begin{figure}[t]
 \includegraphics[scale=0.29]{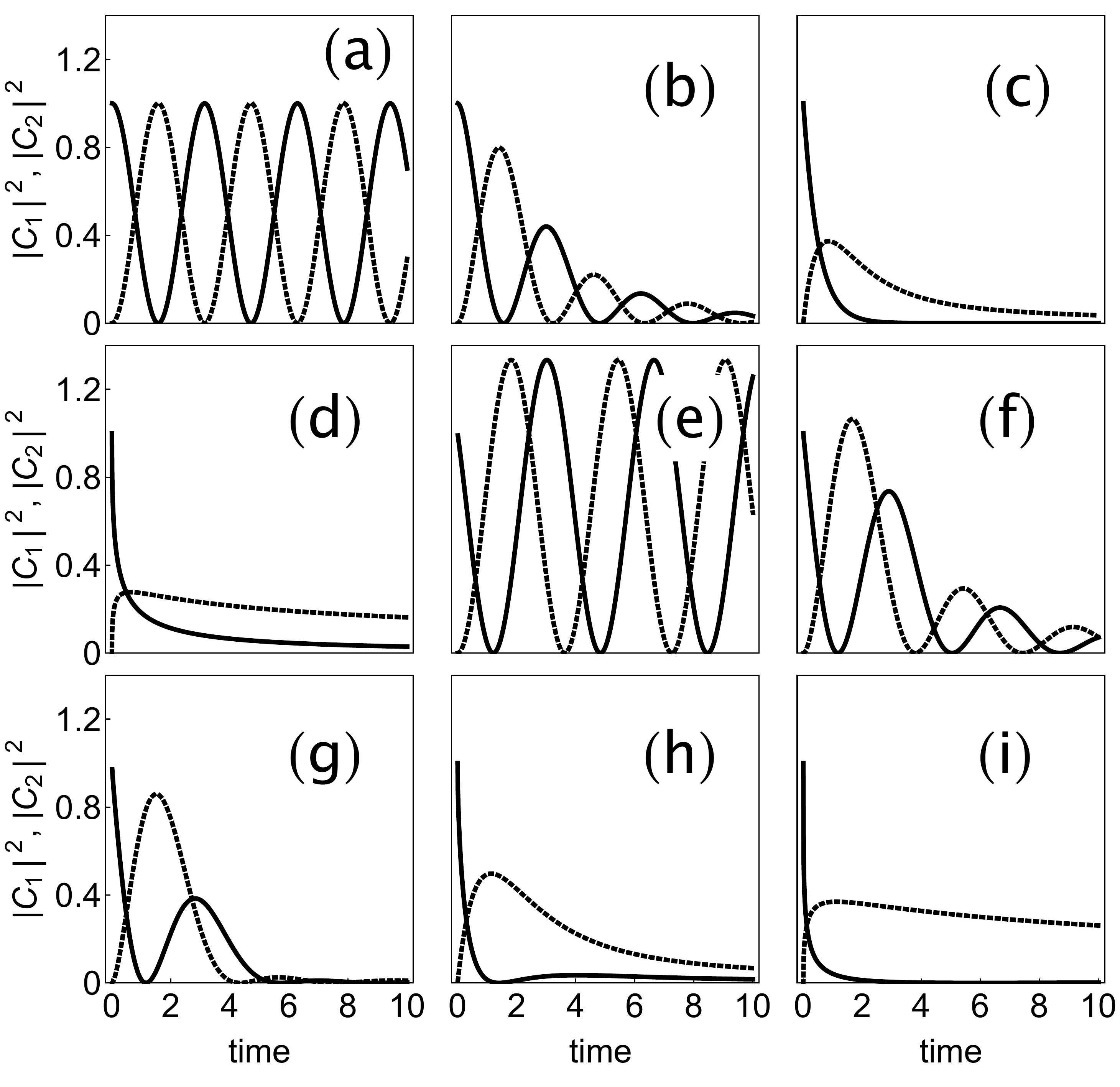}
  \caption{ Dimer amplitudes $|C_{1}(t)|^2$ (continuous line) and $|C_{2}(t)|^2$ (dashed line) for the linear ${\cal{PT}}$ case ($\chi=0, \epsilon\neq 0$) for several fractional derivative orders and various gain/loss parameters. (a) $\alpha=1, \epsilon=0$, (b) $\alpha=0.9, \epsilon=0$, (c) $\alpha=0.5, \epsilon=0$, 
(d) $\alpha=0.25, \epsilon=0$, (e) $\alpha=1, \epsilon=0.5$, (f) $\alpha=0.9,  \epsilon=0.5$, (g) $\alpha=0.8, \epsilon=0.5$, (h) $\alpha=0.5, \epsilon=0.5$, (i) $\alpha=0.25, \epsilon=0.5$.  }
  \label{fig4}
\end{figure}
As we can see, as $\alpha$ decreases from $1$, both amplitudes decrease too, with oscillations that become monotonically decreasing, converging to zero at long  times.

The asymptotic behavior of $C_{1}(t), C_{2}(t)$ depends on the behavior of the Mittag-Leffler functions $E_{\alpha,\beta}(z)$ at large values of 
$|z|$. After writing $z=|z|\exp(i \phi)$, we have\cite{wright}
\be
E_{\alpha,\beta}(z) \approx (1/\alpha)\ Q^{1-\beta}\exp(Q)
\ee	
where $Q=z^{1/\alpha}=\exp((1/\alpha) \log(|z|)+ i\ \phi)$ and $|\phi/\alpha|\leq \pi$ ($\phi=\epsilon^2- V^2 $). This implies,
\be
\exp(Q) = \exp(|z|^{1/\alpha} \cos((1/\alpha)\phi))\times \exp(i\ |z|^{1/\alpha} \sin((1/\alpha)\phi))
\ee
Thus, bounded behavior in time will occur for $(\pi/2)<|\phi/\alpha|<\pi$, while unbounded behavior occurs for $0<|\phi/\alpha|<\pi/2$. In our case, $\phi=\mbox{arg}(\epsilon^2-V^2)=0,\pi$, implying that $C_{1}(t)$ and $C_{2}(t)$ will increase (decrease) asymptotically in time if $(\epsilon^2 - V^2)$ is positive (negative). This behavior is sketched in Fig.4.
\begin{figure}[t]
\includegraphics[scale=0.35,angle=0]{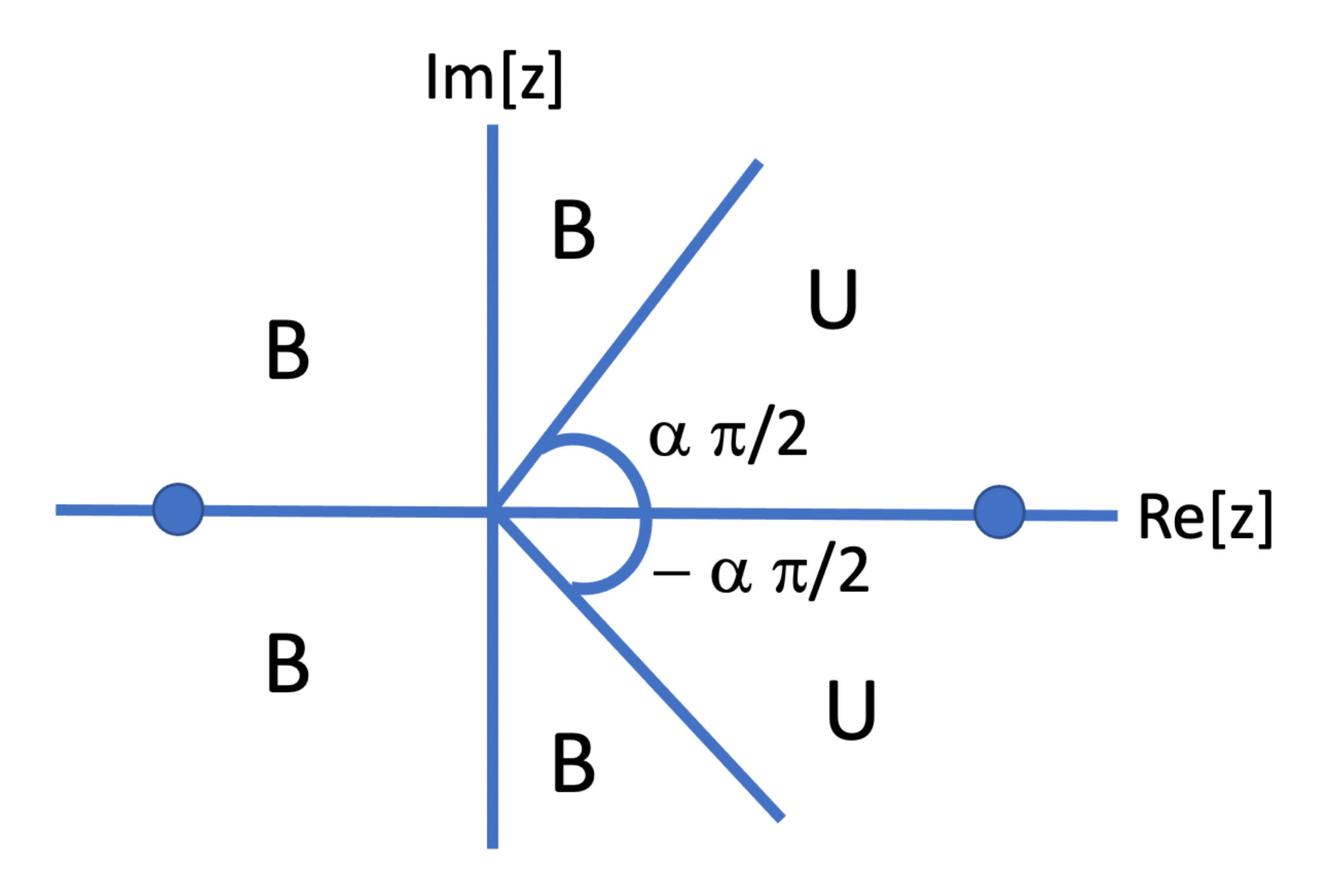}
  \caption{ Asymptotic stability for the amplitudes $C_{1}(t), C_{2}(t)$ for the fractional ${\cal{PT}}$ dimer system (\ref{eq21}). Here, $\mbox {U $\equiv$ unbounded}$, $\mbox{B $\equiv$ bounded}$, and $z=(\epsilon^2 - V^2) t^{2 \alpha}$. The dots denote the position of our two phases, $\mbox{phase}(\epsilon^2-V^2)=0$ for $\epsilon^2-V^2>0$, and $\mbox{phase}(\epsilon^2-V^2)=\pi$ for $\epsilon^2-V^2<0$.
  }
  \label{fig3}
\end{figure}

\subsection{The nonlinear ${\cal{PT}}$ dimer}

We now explore a ${\cal{PT}}$ dimer in the presence of nonlinearity, and subject to fractional evolution equations
\begin{eqnarray}
i {d^{\alpha} C_{1}(t)\over{d t^{\alpha}}} + i \epsilon C_{1}(t) + V C_{2}(t)+ \chi |C_{1}(t)|^2 C_{1}(t)&=&0\nonumber\\
i {d^{\alpha} C_{2}(t)\over{d t^{\alpha}}} - i \epsilon C_{2}(t) + V C_{1}(t) +
\chi |C_{2}(t)|^2 C_{2}(t)&=&0\ \ \ \ \label{eq22}
\end{eqnarray}
In the absence of ${\cal{PT}}$ symmetry ($\epsilon=0$), and for order 
$\alpha=1$, eqs.(\ref{eq22}) have been explored before in the literature\cite{dimer1,dimer2,dimer3}. 
For initial conditions $C_{1}(0)=1, C_{2}(0)=0$, it was shown that they lead to the phenomenon of a seltrapping transition: The existence of a critical nonlinearity parameter $\chi_{c}/V=4$ below which, the long-time average of the square of the amplitudes, $\langle |C_{1,2}|^2 \rangle=(1/T) \int_{0}^{T} |C_{1,2}|^2 dt$ (with $T\gg 1$) is  the same: $\langle |C_{1}|^2 \rangle
= \langle |C_{2}|^2 \rangle=1/2$. At nonlinearity values above $\chi_{c}/V$, $\langle |C_{1}|^2 \rangle$ increases past $1/2$ and converges to $1$ at large $\chi/V$ values, while $\langle |C_{2}|^2 \rangle$ decreases towards zero. The trapped fraction at the initial site, $\langle |C_{1}|^2 \rangle$, increases  abruptly as the critical nonlinearity is crossed.

For a fractional order derivative ($0<\alpha<1$), where we take the Caputo version of the fractional derivative, and in the presence of ${\cal{PT}}$ symmetry, we resort to the Grunwald algorithm\cite{grunwald} 
to compute the time evolution of $C_{1}(t), C_{2}(t)$ for initial conditions $C_{1}(0)=1, C_{2}(0)=0$. This approach is based on finite differences, and in our case leads to the following difference equations:
\begin{eqnarray}
X_{n+1}&=&\sum_{\nu=1}^{n+1} \Phi_{\nu}^{\alpha} X_{n+1-\nu} + i h ( Y_{n}+i  \epsilon X_{n} \nonumber\\
       & &+ \chi\ |X_{n}|^2 X_{n} ) + r_{n+1}^{\alpha} X_{0}\nonumber\\
Y_{n+1}&=&\sum_{\nu=1}^{n+1} \Phi_{\nu}^{\alpha} Y_{n+1-\nu} + i h ( X_{n}-i \epsilon Y_{n} \nonumber\\  
       & & + \chi\ |Y_{n}|^2 Y_{n} ) + r_{n+1}^{\alpha} Y_{0}
\end{eqnarray}
\begin{figure}[t!]
 \includegraphics[scale=0.25]{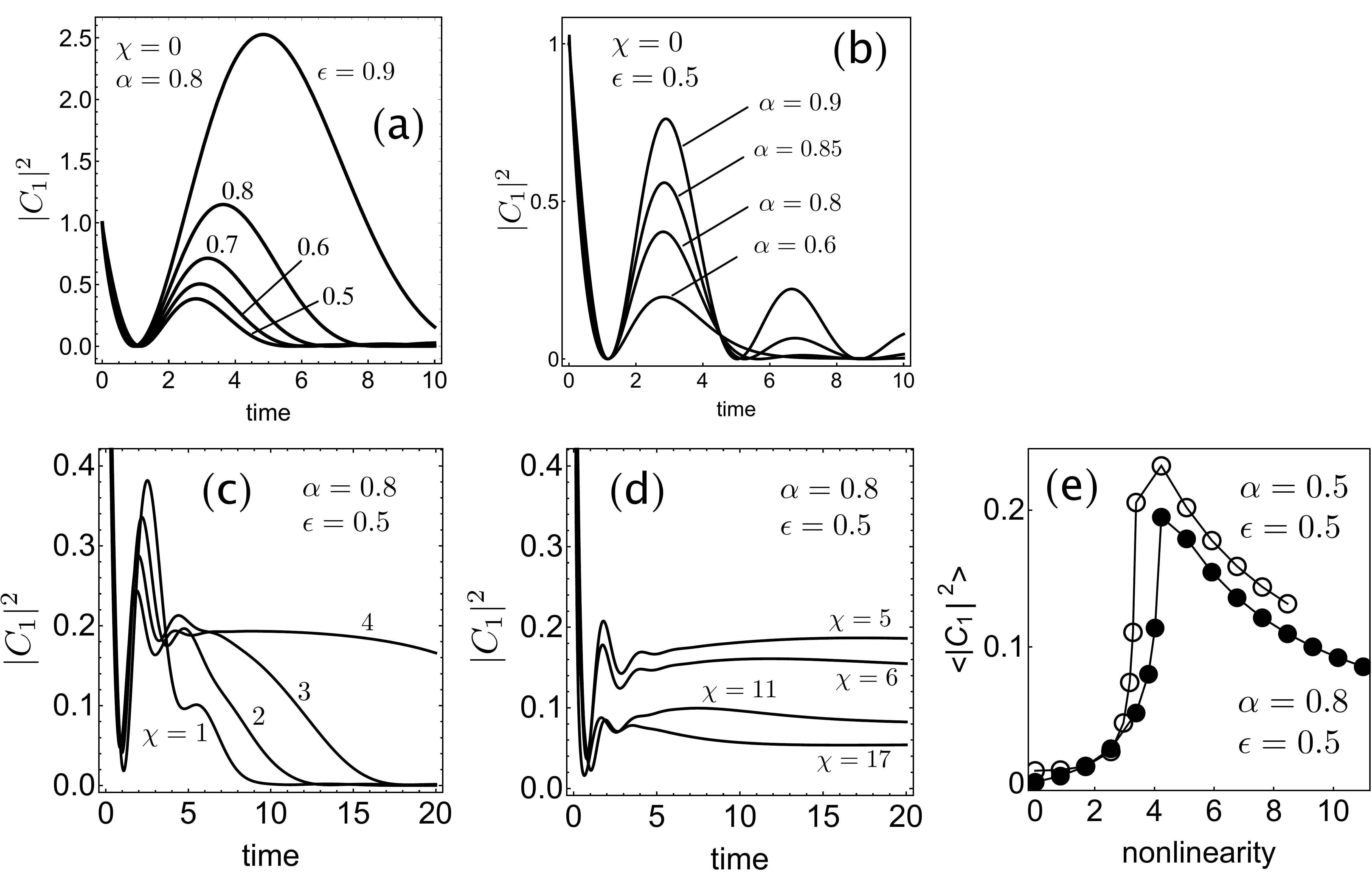}
  \caption{ Dimer amplitude at initial site $|C_{1}(t)|^2$ for the nonlinear ${\cal{PT}}$ case  for several fractional derivative orders $\alpha$, various gain/loss parameters $\epsilon$ and different nonlinearities $\chi$. Lower right: Example of a time-averaged,  trapped fraction at initial site as a function of the nonlinearity parameter ($\epsilon=0.5, \alpha=0.8$).}
  \label{fig5}
\end{figure}
where, $X\equiv C_{1}, Y\equiv C_{2}$, and  
\be
\Phi_{\nu}^{\alpha}=(-1)^{\nu-1} \binom{\alpha}{\nu} \hspace{0.5cm}
r_{\nu}^{\alpha} = {\nu^{-\alpha}\over{\Gamma(1-\alpha)}}.
\ee

Numerical results are shown in Fig.5. In panel (a) we show the behavior of $|C_{1}(t)|^2$ in the linear limit ($\chi=0$), for a fixed $\alpha$ and several different gain/loss parameter values. We see that the effect of increasing $\epsilon$ is to augment the amplitude and decrease the frequency of the oscillation. When $\epsilon$ approaches $V$, the amplitude grows unbounded and the oscillation stops.
In panel (b) we take $\chi=0$ as before, with a fixed gain/loss $\epsilon$ and for several order $\alpha$ values. As we noticed before, the presence of $0<\alpha<1$ induces a decreasing oscillation behavior in $|C_{1}(t)|^2$. If we reduce now the
value of $\alpha$, we see a further decrease of the oscillation amplitude, with little effect on the frequency. We now move to the nonlinear case. In panels (c, d) we show the behavior of $|C_{1}(t)|^2$ in time for fixed $\alpha, \epsilon$ parameters and for several $\chi$ values. Roughly speaking, what we observe here is that there seems to exist a special nonlinearity value below which the curves decrease steadily to zero at long times, and above which they  approach a constant nonzero value in time. To help understand this, we show in Figure 5e $\langle|C_{1}|^2\rangle=(1/T)\int_{0}^{T}|C_{1}|^2 dt$ $(T\gg 1)$, the time-averaged fraction remaining at the initial site vs the nonlinearity strength. As soon as the nonlinearity increases from zero there is a finite amount of trapping at the initial site that increases monotonically with nonlinearity. As the nonlinearity parameter reaches a critical value $\chi_{c}$ whose precise value depends on $\alpha$, there is an abrupt increase in $\langle |C_{1}|^2\rangle$  signaling a seltrapping transition, like in the standard ($\alpha=1$) nonlinear dimer.  
What is interesting though, is that if we continue increasing the nonlinearity past the critical point, the trapped fraction begins to decrease instead of increasing towards unity as in the standard case. This fragility of the trapping could perhaps be a manifestation of the tendency of $\alpha$ to decrease the amplitude of oscillations in general. Thus, what we are seeing here is the interplay of two opposing tendencies: Trapping by nonlinearity and amplitude decay by $\alpha$. 

\section{Conclusions}
We have examined the excitation dynamics in a nonlinear ${\cal{PT}}$ dimer
when the evolution equations are ruled by a fractional-order time derivative, instead of the usual first-order. In the absence of gain/loss and nonlinearity, the effect of the fractional derivative alone is to induce a damping of the oscillatory exchange between the two sites. When ${\cal{PT}}$ symmetry is added, two oscillation regimes are present, a damped oscillatory regime for small gain/loss parameter, and a monotonic growth for large gain/loss parameters. However, there is no stable oscillatory regime, i.e., when there is complete balance between gain and losses, unless $\alpha=1$, the standard case.

Finally, when nonlinearity is added to the picture, we observe  selftrapping at long times at the initial site, that increases steadily as the nonlinearity reaches a critical value of approximately, $\chi\sim 4 V$. Above this threshold, the trapped fraction at the initial site decreases monotonically as nonlinearity is increased further.  This is in marked contrast with the standard case where selftrapping keeps on increasing towards unity with increasing nonlinearity strength.

In spite of these interesting differences with the standard integer derivative case, some  general features remain more or less similar to its standard counterpart: (1) There is exchange between the sites, (2) In the presence of ${\cal{PT}}$ symmetry there are two clearly marked regimes where the amplitudes decrease to zero or diverge to infinity (3) There is still a selftraping transition, whose finer details depend on the value of the fractional derivative order as well as on the strength of the gain/loss parameter. The persistence of this general phenomenology in the face of a different mathematical rate of change, suggests that this phenomenology is robust against ``mathematical perturbations''. This kind of robustness has been also observed for the fractional DNLS equation, where the Laplacian is replaced by a fractional one\cite{luz1,luz2}. Thus, we expect that concepts of power exchange between sites, selftrapping transitions and existence of nonlinear excitations (discrete solitons) to be concepts of wide physical validity for nonlinear discrete systems.

\acknowledgments
This work was supported by Fondecyt Grant 1200120.

\end{document}